\input harvmac
\input epsf
\noblackbox
\newcount\figno
\figno=0
\def\fig#1#2#3{
\par\begingroup\parindent=0pt\leftskip=1cm\rightskip=1cm\parindent=0pt
\baselineskip=11pt
\global\advance\figno by 1
\midinsert
\epsfxsize=#3
\centerline{\epsfbox{#2}}
\vskip 12pt
\centerline{{\bf Figure \the\figno} #1}\par
\endinsert\endgroup\par}
\def\figlabel#1{\xdef#1{\the\figno}}
\def\pano{\par\noindent}

\def\pmb#1{\setbox0=\hbox{#1}%
 \kern-.025em\copy0\kern-\wd0
 \kern.05em\copy0\kern-\wd0
 \kern-.025em\raise.0433em\box0 }
\font\cmss=cmss10
\font\cmsss=cmss10 at 7pt

\def\rlx{\relax\leavevmode}
\def\Cop{\relax\,\hbox{$\inbar\kern-.3em{\rm C}$}}
\def\Rop{\relax{\rm I\kern-.18em R}}
\def\Nop{\relax{\rm I\kern-.18em N}}
\def\Pop{\relax{\rm I\kern-.18em P}}
\def\Zop{\rlx\leavevmode\ifmmode\mathchoice{\hbox{\cmss Z\kern-.4em Z}}
 {\hbox{\cmss Z\kern-.4em Z}}{\lower.9pt\hbox{\cmsss Z\kern-.36em Z}}
 {\lower1.2pt\hbox{\cmsss Z\kern-.36em Z}}\else{\cmss Z\kern-.4em
 Z}\fi}
\def\bphi{{\pmb{$\phi$}}}

\def\H{{\cal H}}

\def\ie{{\it i.e.}}


\lref\polvtwo{J. Polchinski, {\it String Theory, Volume 2, Superstring
Theory and Beyond}, CUP 1998.}

\lref\bgone{O. Bergman, M.R. Gaberdiel, {\it A non-supersymmetric
open string theory and S-duality}, Nucl. Phys.~{\bf B499}, 183 (1997); 
{\tt hep-th/9701137}.}

\lref\klebtse{I.R. Klebanov, A.A. Tseytlin, {\it D-branes and dual
gauge theories in type 0 strings}, Nucl. Phys. {\bf B546}, 155
(1999); {\tt hep-th/9811035}.}

\lref\bgfour{O. Bergman, M.R. Gaberdiel, {\it Dualities of Type 0
strings}, JHEP {\bf 9907}, 022 (1999); {\tt hep-th/9906055}.}

\lref\biasag{M. Bianchi, A. Sagnotti, {\it On the systematics of
open string theories}, Phys. Lett. {\bf B247}, 517 (1990);  
{\it Twist symmetry and open string Wilson lines}, Nucl. Phys. 
{\bf B361}, 519 (1991).}  

\lref\sensix{ A. Sen, {\it BPS D-branes on non-supersymmetric
cycles}, JHEP {\bf 9812}, 021 (1998); {\tt hep-th/9812031}.}

\lref\bgthree{O. Bergman, M.R. Gaberdiel, {\it  Non-BPS states in
Heterotic -- Type IIA duality}, JHEP {\bf 9903}, 013 (1999); 
{\tt hep-th/9901014}.}

\lref\ddg{D-E. Diaconescu, M.R. Douglas, J. Gomis, {\it Fractional
branes and wrapped branes}, JHEP {\bf 9802}, 013 (1998); 
{\tt hep-th/9712230}.}

\lref\gabsen{M.R. Gaberdiel, A. Sen, {\it Non-supersymmetric
D-Brane configurations with Bose-Fermi degenerate open string
spectrum}, JHEP {\bf 9911}, 008 (1999); {\tt hep-th/9908060}.}

\lref\recksch{A. Recknagel, V. Schomerus, {\it D-branes in Gepner
models}, Nucl. Phys. {\bf B531}, 185 (1998), {\tt hep-th/9712186}.}

\lref\gutsat{M. Gutperle, Y. Satoh, {\it D0-branes in Gepner models
and N=2 black holes}, Nucl. Phys. {\bf B555}, 477 (1999); 
{\tt hep-th/9902120}.}

\lref\bdlr{I. Brunner, M.R. Douglas, A. Lawrence, C. R\"omelsberger,
{\it D-branes on the quintic}, {\tt hep-th/9906200}.}

\lref\dioro{D.-E. Diaconescu, C. R\"omelsberger, {\it D-Branes and
bundles on elliptic fibrations}, Nucl. Phys. {\bf B574}, 245 (2000);
{\tt hep-th/9910172}.}

\lref\kllw{P. Kaste, W. Lerche, C. A. Lutken, J. Walcher, 
{\it D-branes on K3-fibrations}, {\tt hep-th/9912147}.}

\lref\scheid{E. Scheidegger, {\it D-branes on some one- and
two-parameter Calabi-Yau hypersurfaces}, JHEP {\bf 0004}, 003 (2000);
{\tt hep-th/9912188}.}

\lref\brusch{I. Brunner, V. Schomerus, {\it D-branes at singular
curves of Calabi-Yau compactifications}, JHEP {\bf 0004}, 020 (2000);
{\tt hep-th/0001132}.}

\lref\nano{M. Naka, M. Nozaki, {\it Boundary states in Gepner models},
JHEP {\bf 0005}, 027 (2000); {\tt hep-th/0001037}.}

\lref\dfr{M.R. Douglas, B. Fiol, C. R\"omelsberger, {\it The spectrum
of BPS branes on a noncompact Calabi-Yau}, {\tt hep-th/0003263}.}

\lref\diadoug{D.-E. Diaconescu, M.R. Douglas, {\it D-branes on stringy
Calabi-Yau manifolds}, {\tt hep-th/0006224}.}

\lref\pssa{G. Pradisi, A. Sagnotti, Ya. S. Stanev, {\it Planar
duality in $SU(2)$ WZW models}, Phys. Lett. {\bf B354}, 279 (1995);
{\tt hep-th/9503207}.}

\lref\pssb{G. Pradisi, A. Sagnotti, Ya. S. Stanev, {\it The open
descendants of non-diagonal $SU(2)$ WZW models}, Phys. Lett. {\bf B356},
230 (1995); {\tt hep-th/9506014}.}

\lref\aleksch{A. Alekseev, V. Schomerus, {\it D-branes in the WZW
model}, Phys. Rev. {\bf D60}, 061901 (1999); {\tt hep-th/9812193}.}

\lref\ars{A. Alekseev, A. Recknagel, V. Schomerus, 
{\it Non-commutative world-volume geometries: branes on $SU(2)$ and
fuzzy spheres},  JHEP {\bf 9909}, 023 (1999); {\tt hep-th/9908040}.}

\lref\fffs{G. Felder, J. Fr\"ohlich, J. Fuchs, Ch. Schweigert, 
{\it The geometry of WZW branes}, J. Geom. Phys. {\bf 34}, 162 (2000);
{\tt hep-th/9909030}.}

\lref\hnt{Y. Hikida, M. Nozaki, T. Takayanagi, {\it Tachyon
condensation on fuzzy sphere and noncommutative solitons}, 
{\tt hep-th/0008023}. }

\lref\vafawitten{C. Vafa, E. Witten, {\it On orbifolds with discrete
torsion}, J. Geom. Phys. {\bf 15}, 189 (1995); {\tt hep-th/9409188}.}

\lref\stef{B. Stefa\'nski, {\it Dirichlet Branes on a Calabi-Yau 
three-fold orbifold}, {\tt hep-th/0005153}.}

\lref\bgfive{O. Bergman, M.R. Gaberdiel, {\it Consistency of
orbifolds}, Phys. Lett. {\bf B481}, 379 (2000); {\tt hep-th/0001130}.}

\lref\vafa{C. Vafa, {\it Modular invariance and discrete torsion on
orbifolds}, Nucl. Phys. {\bf B273}, 592 (1986).}

\lref\orbifolds{L. Dixon, J. Harvey, C. Vafa, E. Witten, {\it Strings
on orbifolds I and II}, Nucl. Phys. {\bf B261}, 678 (1985) and 
Nucl. Phys. {\bf B274}, 295 (1986).}

\lref\gomis{J. Gomis, {\it D-branes on orbifolds with discrete torsion
and topological obstruction}, JHEP {\bf 0005}, 006 (2000); 
{\tt hep-th/0001200}.}

\lref\douglas{M.R. Douglas, {\it D-branes and discrete torsion}, {\tt
hep-th/9807235}.}

\lref\dougfiol{M.R. Douglas, B. Fiol, {\it  D-branes and discrete
torsion II}, {\tt hep-th/9903031}.}

\lref\lerdarev{A. Lerda, R. Russo, {\it Stable non-BPS states 
in string theory: a pedagogical review},  Int. J. Mod. Phys. 
{\bf A15}, 771 (2000); {\tt hep-th/9905006}.} 

\lref\mrgrev{M.R. Gaberdiel, {\it Lectures on Non-BPS Dirichlet
branes}, {\tt hep-th/0005029}.}

\lref\gabste{M.R. Gaberdiel, B. Stefa\'nski, {\it Dirichlet
branes on orbifolds}, Nucl. Phys. {\bf B578}, 58 (2000); 
{\tt hep-th/9910109}.} 

\lref\sena{A. Sen, {\it Stable non-BPS bound states of BPS
D-branes}, JHEP {\bf 9808}, 010 (1998); {\tt hep-th/9805019}.}

\lref\bgtwo{O. Bergman, M.R. Gaberdiel, {\it Stable non-BPS
D-particles}, Phys. Lett.~{\bf B441}, 133 (1998);
{\tt hep-th/9806155}.}

\lref\polcai{J. Polchinski, Y. Cai, {\it Consistency of open
superstring theories}, Nucl. Phys. {\bf B296}, 91 (1988).}

\lref\clny{C. Callan, C. Lovelace, C. Nappi, S. Yost, {\it Loop
corrections to superstring equations of motion}, Nucl. Phys. 
{\bf B308}, 221 (1988).} 

\lref\VecLic{P. Di~Vecchia, A. Liccardo, {\it D branes in string
theory, I \& II}, {\tt hep-th/9912161} and {\tt hep-th/9912275}.}

\lref\polch{J. Polchinski, {\it Tensors from K3 orientifolds}, 
Phys. Rev. {\bf D55}, 6423 (1997); {\tt hep-th/9606165}.}

\lref\mukray{S. Mukhopadhyay, K. Ray, {\it D-branes on fourfolds with
discrete torsion},  Nucl. Phys. {\bf B576}, 152 (2000); 
{\tt hep-th/9909107}.}

\lref\berlei{D. Berenstein, R.G. Leigh, {\it Discrete torsion, AdS/CFT
and duality}, JHEP {\bf 0001}, 038 (2000); {\tt hep-th/0001055}.}

\lref\diagom{D.-E. Diaconescu, J. Gomis, {\it Fractional branes and
boundary states in orbifold theories}, {\tt hep-th/9906242}.}

\lref\kleinraba{M. Klein, R. Rabadan, {\it Orientifolds with discrete
torsion}, JHEP {\bf 0007}, 040 (2000); {\tt hep-th/0002103}.}

\lref\hankol{A. Hanany, B. Kol, {\it  On orientifolds, discrete
torsion, branes and M theory}, JHEP {\bf 0006}, 013 (2000);
{\tt hep-th/0003025}.}

\lref\aspmor{P.S. Aspinwall, D.R. Morrison, {\it Stable singularities
in string theory}, Commun. Math. Phys. {\bf 178}, 115 (1996);
{\tt hep-th/9503208}.}

\lref\sharpe{E.R. Sharpe, {\it Discrete torsion and gerbes I \& II}, 
{\tt hep-th/9909108} and {\tt 9909120}.}

\lref\sharpetwo{E.R. Sharpe, {\it Discrete torsion}, 
{\tt hep-th/0008154}.}

\lref\dougmoore{M.R. Douglas, G. Moore, {\it D-branes, quivers and ALE
instantons}, {\tt hep-th/9603167}.}

\lref\cragab{B. Craps, M.R. Gaberdiel, {\it in preparation}.}

\lref\aadds{C. Angelantonj, I. Antoniadis, G. D'Appollonio, E. Dudas,
A. Sagnotti, {\it Type I vacua with brane supersymmetry breaking}, 
Nucl.Phys. {\bf B572}, 36 (2000); {\tt hep-th/9911081}.}

\Title{\vbox{
\hbox{hep--th/0008230}
\hbox{DAMTP-2000-89}}}
{\vbox{\centerline{Discrete torsion orbifolds and D-branes}
}}
\centerline{Matthias R. Gaberdiel\footnote{$^{\star}$}{\tt
e-mail: M.R.Gaberdiel@damtp.cam.ac.uk, 
mrg@mth.kcl.ac.uk}}\footnote{}{Address after 1 
October 2000: Department of Mathematics, King's College London,
Strand, London WC2R 2LS, U.K.}  
\bigskip
\centerline{\it Department of Applied Mathematics and Theoretical
Physics} 
\centerline{\it University of Cambridge}
\centerline{\it Centre for Mathematical Sciences}
\centerline{\it Wilberforce Road, Cambridge CB3 0WA, U.K.}
\smallskip
\vskip2cm
\centerline{\bf Abstract}
\bigskip
\noindent
D-branes on orbifolds with and without discrete torsion are analysed
in a unified way using the boundary state formalism. For the example
of the $\Zop_2\times\Zop_2$ orbifold it is found that both the theory
with and without discrete torsion possess D-branes whose world-volume
carries conventional and projective representations of the orbifold
group. The resulting D-brane spectrum is shown to be consistent with
T-duality.

\bigskip

\Date{Revised 10/2000}

\newsec{Introduction}

Much has been learned in the last few years about D-branes in string
theory, and this has deepened our understanding of many aspects of the
theory. D-branes play a crucial r\^ole in testing the various duality
relations between string theories. D-branes also provide new insights
into the background geometry of string theory since the geometry can
be analysed in terms of the low-energy theory on the world-volume of a
brane probe.   

The D-brane spectrum of a number of theories is understood in
detail. These include the standard ten-dimensional Type IIA, IIB and I
theory (see \refs{\polvtwo} for a review), as well as their
non-supersymmetric cousins, Type 0A, 0B and 0
\refs{\bgone,\klebtse,\bgfour} (for an earlier non-supersymmetric
orientifold construction see also \refs{\biasag}). It is understood how
the D-brane spectrum is modified upon compactification on
supersymmetric orbifolds and near ALE singularities  
\refs{\dougmoore,\sensix,\bgthree,\ddg,\diagom,\gabsen,\gabste}. 
There has also been progress in understanding the D-brane spectrum of
Gepner models \refs{\recksch,\gutsat}, more general Calabi-Yau
manifolds
\refs{\bdlr,\dioro,\kllw,\scheid,\brusch,\nano,\dfr,\stef,\diadoug}
and WZW models \refs{\pssa,\pssb,\aleksch,\ars,\fffs,\hnt}.  

Recently, D-branes on orbifolds and orientifolds with discrete torsion
\refs{\vafa,\vafawitten} have attracted some interest 
\refs{\douglas,\dougfiol,\diagom,\mukray,\aadds,\berlei,\gomis,
\kleinraba,\hankol}. The geometry of discrete torsion orbifolds is
only partially understood
\refs{\vafawitten,\aspmor,\sharpe,\sharpetwo}, and it is therefore
interesting to study what can be learned about it from the analysis of
D-brane probes. It was argued in \refs{\douglas} that D-branes in
orbifolds with discrete torsion are characterised by the property that
the representation of the orbifold group in the corresponding open
string description is a {\it projective} representation. One such
brane was analysed in detail, and it was found that its moduli space
has a structure that is in agreement with the expectations based on
\refs{\vafawitten,\aspmor}. This analysis was extended in
\refs{\dougfiol} to a more general class of orbifolds. 

As was mentioned in \refs{\dougfiol}, one of the models is T-dual to
an orbifold without discrete torsion \refs{\vafawitten}. Since the
orbifold with discrete torsion has a brane for which the orbifold
group acts projectively on the Chan-Paton factors of the open string,
this raises the question of what the T-dual of this brane in the
theory without discrete torsion should be 
\refs{\dougfiol}. (Similarly, it also raises the question of what the
images of the fractional branes of the theory without discrete torsion
in the theory with discrete torsion are.) In this paper we propose an
answer to both of these questions: we shall argue that {\it both}
orbifolds with and without discrete torsion have D-branes for which
the open string has a projective representation of the orbifold group;
conversely, {\it both} orbifolds also have D-branes for which the open
string has a proper representation of the orbifold group. Our analysis
is based on the construction of boundary states that can be performed
irrespective of whether the theory has discrete torsion or not. In
fact, since discrete torsion is a relative concept, there does not
seem to exist an abstract sense in which a given theory has discrete
torsion or not; one should therefore expect that the standard analysis
for branes on orbifolds without discrete torsion should equally apply
for the case with discrete torsion. The D-brane spectrum we find is
consistent with T-duality, it accounts for all the R-R charges of the
theory, and it leads to open strings that satisfy the open-closed
consistency condition.
\vskip 0.3cm

The paper is organised as follows. In section~2 we review briefly the
main concepts of discrete torsion, and in section~3 we describe a
$\Zop_2\times\Zop_2$ orbifold with and without discrete torsion in
detail. The boundary states for both orbifolds are constructed in
section~4. In particular, we describe in some detail the boundary
state description of the D-brane that leads to a projective
representation of the orbifold group in the open string. We also
explain why T-duality requires that orbifolds with discrete torsion
also have branes that lead to conventional representations in the open
string, and why, conversely, orbifolds without discrete torsion also
have to have branes with projective open string representations. In
section~5 we re-examine the consistency argument of Gomis
\refs{\gomis} and explain why it is consistent with what we
propose. Finally section~6 contains some conclusions. We have included
an appendix where the relation between discrete torsion phases and the
second cohomology of the orbifold group with coefficients in $U(1)$ is
spelled out in some detail.

\newsec{Discrete torsion}

Let us briefly recall the definition of discrete torsion in
orbifolds. Suppose we consider the orbifold of a closed string theory 
on ${\cal M}$ by the (abelian) group $\Gamma$. As is well known
\refs{\orbifolds}, the orbifold theory consists of the invariant
subspace of the original theory under the action of the orbifold
group $\Gamma$. In addition, the theory has so-called twisted sectors
that describe those closed strings that are only closed in 
${\cal M}/\Gamma$, but not in ${\cal M}$. For an abelian orbifold we
have a twisted sector $\H_h$ for each element $h\in\Gamma$. Each
twisted sector has to be projected again onto the states that are
invariant under the orbifold group $\Gamma$; the corresponding
projector is of the form 
\eqn\proj{ P = {1 \over |\Gamma|} \sum_{g\in\Gamma} g \,.}
The total partition function of the theory is then 
\eqn\part{ Z(q,\bar{q}) 
= {1 \over |\Gamma|}  \sum_{g,h\in\Gamma} Z(q,\bar{q};g,h)\,,}
where 
\eqn\Zdef{Z(q,\bar{q};g,h) = 
\hbox{Tr}_{\H_h}(q^{L_0} \bar{q}^{\bar{L}_0} g) \,.}

The theory with discrete torsion \refs{\vafa} is characterised by the
property that the partition function is 
\eqn\torsion{ Z(q,\bar{q}) = {1 \over |\Gamma|}  
\sum_{g,h\in\Gamma} \epsilon(g,h) Z(q,\bar{q};g,h)\,,}
where $\epsilon(g,h)$ are phases. Modular invariance at one loop
requires that 
\eqn\consistone{ \epsilon(g,h) = \epsilon(g^a h^b,g^c h^d) \qquad
\hbox{where $ad-bc=1$ and $a,b,c,d\in\Zop$.}}
Furthermore modular invariance on higher genus surfaces, together
with the factorization property of loop amplitudes, implies that the  
$\epsilon(g,h)$ have to define a one-dimensional representation of
$\Gamma$, 
\eqn\consisttwo{\epsilon(g_1 g_2,h)= 
           \epsilon(g_1,h) \epsilon(g_2,h)\,.} 

The set of inequivalent different torsion theories are classified by
the second cohomology group of $\Gamma$ with values in $U(1)$,
$H^2(\Gamma,U(1))$.\footnote{$^\dagger$}{The following discussion
follows closely \refs{\gomis}.} This cohomology group consists of the
two-cocycles $c(g,h)\in U(1)$ satisfying the cocycle condition 
\eqn\cocycle{c(g_1,g_2 g_3) c(g_2,g_3) = c(g_1g_2,g_3) c(g_1,g_2)\,,}
where we identify cocycles that differ by a coboundary, 
\eqn\cobound{ c'(g,h) = {c_g c_h \over c_{gh}} c(g,h)\,.}
(Here $c_g\in U(1)$ for each $g\in\Gamma$.) Indeed, for each such
cocycle one can define
\eqn\eps{ \epsilon(g,h) = {c(g,h) \over c(h,g)} \,.}
It follows immediately from this definition that 
\eqn\pro{ \epsilon(g,g) = 1 \qquad \epsilon(g,h) =
\epsilon(h,g)^{-1}\,,} 
and a short calculation shows that \cocycle\ implies
\consisttwo. Together with \pro\ this is then sufficient to prove
\consistone\ (see \refs{\vafa}). It is also manifest from \eps\ that
the definition of $\epsilon(g,h)$ is the same for cocycles that differ
by a coboundary. Conversely, one can construct a cocycle $c$ for each
consistent set of discrete torsion phases so that \eps\ is satisfied;
this is described in detail in the
appendix.\footnote{$^\ddagger$}{Naively one may think that $c(g,h)$
can simply be defined by $c(g,h)=\epsilon(g,h)^{1/2}$ since
\consistone\ and \consisttwo\ imply that
$\epsilon(g,h)=\epsilon(h,g)^{-1}$, and therefore \eps\ reproduces
$\epsilon(g,h)$. In addition \consisttwo\ implies that the square
of the left hand side in \cocycle\ equals the square of the right hand
side, but this only implies that \cocycle\ holds up to sign. In fact,
there does not seem to exist a `natural' sign conventions for the
definition of $c(g,h)=\epsilon(g,h)^{1/2}$ that lead to $c(g,h)$
satisfying \cocycle. In the appendix we shall therefore follow a
different route.} 

Because of the relation between \part\ and \proj, the modification of
the partition function implies that the projection operator onto
physical states in the sector $\H_h$ is modified to be  
\eqn\projmod{ P|_{\H_h} = {1 \over |\Gamma|} 
\sum_{g\in\Gamma} \epsilon(g,h) g|_{\H_h} \,.}
In particular, this implies that a physical state in the twisted
sector $\H_h$ satisfies
\eqn\twisted{g |s\rangle_{\H_h} = \epsilon(g,h)^\ast |s\rangle_{\H_h}
\,.}
Alternatively, one can interpret the theory with discrete torsion as
the theory where $g\in\Gamma$ acts on the $\H_h$ sector as 
\eqn\torsiontwo{\hat{g}|_{\H_h} = \epsilon(g,h) g|_{\H_h} \,.}
Because of \consisttwo\ this gives a well-defined action of $\Gamma$
on each $\H_h$. From the point of view of conformal field theory, the
orbifold with discrete torsion can therefore be thought of as a
standard orbifold (without discrete torsion) where the elements of
$\Gamma$ act as $\hat{g}$ on the various sectors. In particular, this
implies that (at least from this perspective) there is no abstract
sense in which one can say that a given orbifold is an orbifold with
discrete torsion; rather, discrete torsion is a relative concept that
describes how to obtain a consistent orbifold from another consistent
orbifold (in a way that does not modify the action of the orbifold
group in the untwisted sector). One should therefore expect that
D-branes on orbifolds `with torsion' can be discussed and described
using the same techniques as in the case `without torsion'. This is
indeed what we shall find.

\newsec{An example of an orbifold with discrete torsion}
\pano

The simplest example of an orbifold where discrete torsion is possible
is the $\Zop_2\times\Zop_2$ orbifold of Type II on $T^6$, where 
the generators of the orbifold group, $g_1$ and $g_2$ act as 
\vskip 0.4cm
\vbox{
\centerline{\vbox{
\hbox{\vbox{\offinterlineskip
\def\tablespace{height2pt&\omit&&\omit&&\omit&&\omit&&\omit&&\omit&&
 \omit&\cr}
\def\tablerule{\tablespace\noalign{\hrule}\tablespace}
\def\tableruleA{\tablespace\noalign{\hrule height1pt}\tablespace}
\hrule\halign{&\vrule#&\strut\hskip0.2cm\hfil#\hfill\hskip0.2cm\cr
\tablespace
& && $x^3$ && $x^4$  && $x^5$ && $x^6$ && $x^7$ && $x^8$ &\cr
\tableruleA
& $g_1$ && $+$ && $+$ && $-$ && $-$ && $-$ && $-$ &\cr \tablespace
& $g_2$ && $-$ && $-$ && $+$ && $+$ && $-$ && $-$ &\cr \tablespace
& $g_3$ && $-$ && $-$ && $-$ && $-$ && $+$ && $+$ &\cr 
\tablespace}\hrule}}}}
\centerline{
\hbox{{\bf Table 1:} {\it The action of the orbifold group.}}}}
\vskip 0.5cm
\noindent
Here we have defined $g_3=g_1 g_2$; a $+$ sign in the above table
means that $g_i$ acts as the identity on the corresponding coordinate,
whereas a $-$ sign indicates that $g_i$ acts as $x^j\mapsto - x^j$.

In this case, there are two possible choices for the signs
$\epsilon(g,h)$: either we choose $\epsilon(g,h)=+1$ for all
$g,h\in\Gamma\cong \Zop_2\times \Zop_2$, or we define
\eqn\tor{ \epsilon(g,h) = \left\{ 
\eqalign{ +1 \quad & \hbox{if $g=e$, $h=e$ or $g=h$} \cr
          -1 \quad & \hbox{otherwise.}} \right.}
The first solution corresponds to the trivial cocycle $c(g,h)=+1$ for
all $g,h\in\Gamma$, whereas \tor\ comes from
\eqn\torc{ c(g,h) = \left\{ 
\eqalign{ +1 \quad & \hbox{if $g=e$ or $h=e$ or $g=g_2$ or $h=g_1$} \cr
          -1 \quad & \hbox{otherwise.}} \right.}

In either case, the resulting manifold is a Calabi-Yau manifold
\refs{\vafawitten}, and the theory therefore preserves
supersymmetry. This can be used to determine the relevant
GSO-projections in the various sectors. For the case of the Type IIA
orbifold, the correct GSO-projection is 
\eqn\GSOA{{\bf IIA} \qquad
\eqalign{ {1\over 4} (1+(-1)^F) (1+(-1)^{\tilde{F}}) \qquad
& \hbox{in all NS-NS sectors} \cr
{1\over 4} (1+(-1)^F) (1-(-1)^{\tilde{F}}) \qquad & 
\hbox{in all R-R sectors,}}}
while for the case of the Type IIB orbifold we have
\eqn\GSOB{{\bf IIB} \qquad
\eqalign{ {1\over 4} (1+(-1)^F) (1+(-1)^{\tilde{F}}) \qquad
& \hbox{in all NS-NS sectors} \cr
{1\over 4} (1+(-1)^F) (1+(-1)^{\tilde{F}}) \qquad & 
\hbox{in all R-R sectors.}}}
Alternatively, this can be determined using the approach of
\refs{\bgfive} (see further below). The GSO-projection is the same for
either choice of $\epsilon$. 

The definition of $g_i$ in all sectors that have fermionic zero modes
is {\it a priori} ambiguous; we choose the convention that on the
ground states $g_1$ acts as\footnote{$^\star$}{The fermionic zero
modes are normalised so that
$\{\psi^\mu_0,\psi^\nu_0\}=\delta^{\mu,\nu}$ and similarly for 
$\widetilde\psi^\mu_0$.}
\eqn\orbone{ g_1 = \prod_{i\in{\cal Z}} \left(\sqrt{2} \psi^i_0\right)
\prod_{i\in{\cal Z}} \left(\sqrt{2} \widetilde\psi^i_0\right) \,,}
where ${\cal Z}\subset \{5,6,7,8\}$ is the set of coordinates along
which there are fermionic zero modes, and the ordering of the zero
modes in the two products is the same. The formulae for $g_2$ and
$g_3$ are analogous. There exists another natural definition for
$g_i$, where $g_1$ acts on the ground states as 
\eqn\orbonep{ \hat{g}_1 = \prod_{i\in{\cal Z}} \left( 2 \psi^i_0
\widetilde\psi^i_0\right) \,,}
and $\hat{g}_2$ and $\hat{g}_3$ are analogously defined. These two
definitions are precisely related by discrete torsion, namely 
\eqn\torsion{\hat{g}_i |_{\H_{g_j}} 
     = \epsilon(g_i,g_j) g_i|_{\H_{g_j}}\,,} 
where $\epsilon$ is defined as in \tor. In this example it is clear
that it is a matter of convention which of the two theories one
interprets as the orbifold with torsion, and which as the orbifold
without torsion. 

However, there is one statement that can be made irrespective of this 
convention: the T-dual of the IIA theory without torsion (where we
T-dualise along $x^3,x^5$ and $x^7$, say) is the IIB theory with
torsion and vice versa \refs{\vafawitten}. In fact, the theory where
$g_i$ acts as in \orbone\ has the Hodge diamond
\eqn\hodge{\left(\matrix{ 1&0&0&1\cr 0&3&51&0 \cr
0&51&3&0 \cr 1 & 0 & 0 & 1}\right) \,,}
while for the theory where $g_i$ acts as in \orbonep\ the Hodge
diamond is 
\eqn\hodged{\left(\matrix{ 1&0&0&1\cr 0&51&3&0 \cr
0&3&51&0 \cr 1 & 0 & 0 & 1}\right) \,.}
This suggests that the two theories are actually mirror partners of
each other which is indeed the case \refs{\vafawitten}. Since 
we know how D-branes behave under T-duality, this example provides
consistency conditions on the D-brane spectrum of theories with and
without discrete torsion. We shall check later that our description of
the D-brane spectrum satisfies this constraint.

\newsec{Boundary states and D-branes}
\pano

One method for the analysis of D-branes is the boundary state
approach in which D-branes are described in terms of coherent
(boundary) states of the underlying closed string theory  
\refs{\polcai,\clny,\bgone} (see also \refs{\lerdarev,\VecLic,\mrgrev}
for reviews). For the case of the above orbifold (without discrete
torsion), this analysis has been performed in \refs{\stef}, and we
collect the relevant results here. Following \refs{\gabste,\stef}, we
denote a Dirichlet $p$-brane as $(r;s_1,s_2,s_3)$ where
$p=r+s_1+s_2+s_3$, provided that it has $r+1$ Neumann boundary
conditions along the directions that have not been affected by the
orbifold, \ie\ $x^0,x^1,x^2,x^9$, and $s_i$ Neumann boundary
conditions along the directions $x^{2i+1}$ and $x^{2i+2}$.  

As is familiar in theories with world-sheet fermions, the actual
boundary state is a linear combination of boundary states for
different values of the parameter $\eta$ labelling the different
spin structures. In each sector of the theory, there is a (up to
normalisation) unique linear combination that is invariant under
$(-1)^{F}$; under the action of the various other operators (where the
action of $g_i$ is defined as in \orbone), this state transforms as
follows:
\vskip 0.4cm
\centerline{\vbox{
\hbox{\vbox{\offinterlineskip
\def\tablespace{height2pt&\omit&&\omit&&\omit&&\omit&&\omit&\cr}
\def\tablerule{\tablespace\noalign{\hrule}\tablespace}
\def\tableruleA{\tablespace\noalign{\hrule height1pt}\tablespace}
\hrule\halign{&\vrule#&\strut\hskip0.2cm\hfil#\hfill\hskip0.2cm\cr
\tablespace
& {} && $(-1)^{\tilde{F}}$ && $g_1$  && $g_2$ && $g_3$ &\cr
\tableruleA
& NS-NS;U && $+1$ && $+1$ && $+1$ && $+1$ &\cr \tablespace
& R-R;U && $(-1)^{r+s_1+s_2+s_3+1}$ && $(-1)^{s_2+s_3}$ 
&& $(-1)^{s_1+s_3}$ && $(-1)^{s_1+s_2}$ &\cr \tablerule
& NS-NS;T$_{g_1}$ && $(-1)^{s_2+s_3}$ && $(-1)^{s_2+s_3}$ 
&& $(-1)^{s_3}$ && $(-1)^{s_2}$ &\cr \tablespace
& R-R;T$_{g_1}$ && $(-1)^{r+s_1+1}$ && $+1$ 
&& $(-1)^{s_1}$ && $(-1)^{s_1}$ &\cr \tablerule
& NS-NS;T$_{g_2}$ && $(-1)^{s_1+s_3}$ && $(-1)^{s_3}$ 
&& $(-1)^{s_1+s_3}$ && $(-1)^{s_1}$ &\cr \tablespace
& R-R;T$_{g_2}$ && $(-1)^{r+s_2+1}$ && $(-1)^{s_2}$ 
&& $+1$ && $(-1)^{s_2}$ &\cr \tablerule
& NS-NS;T$_{g_3}$ && $(-1)^{s_1+s_2}$ && $(-1)^{s_2}$ 
&& $(-1)^{s_1}$ && $(-1)^{s_1+s_2}$ &\cr \tablespace
& R-R;T$_{g_3}$ && $(-1)^{r+s_3+1}$ && $(-1)^{s_3}$ 
&& $(-1)^{s_3}$ && $+1$ &\cr \tablespace
\tablespace}\hrule}}}}
\centerline{
\hbox{{\bf Table 2:} {\it The transformation properties of the
boundary states $(r;s_1,s_2,s_3)$.}}}
\vskip 0.5cm
\noindent

\subsec{The theory without discrete torsion}

Having collected this information, we can now describe the D-brane
spectrum of these orbifold theories. Let us first consider the case
`without torsion' (\ie\ the theory where $g_i$ acts as \orbone, rather
than as \orbonep\ in the twisted sectors). We shall mainly concentrate
on the BPS branes, although these theories also contain interesting
(stable) non-BPS branes \refs{\sena,\bgtwo,\stef}. First of all the
orbifold theory contains the familiar `fractional branes'; these
branes are localised at one of the $64$ fixed planes of $\Gamma$, and
their open string spectrum (of the open string that begins and ends on
the same brane) is of the form 
\eqn\open{ {1\over 8} (1+(-1)^F) (1+ g_1) (1+g_2)\,.}
The corresponding boundary state has a non-trivial component in all
untwisted and twisted sectors (where the twisted sectors are all
associated to the same fixed plane). In order for this to be possible,
we have the following restriction on $(r;s_1,s_2,s_3)$:
\eqn\fractional{ \hbox{Fractional D-branes}  \qquad \qquad
\eqalign{\hbox{\bf IIA:} \quad & \hbox{$r$ even and $s_i$ even,} \cr
\hbox{\bf IIB:} \quad & \hbox{$r$ odd and $s_i$ even.}}}
Unlike what happens in simpler orbifold theories (such as the ones
studied in \refs{\bgtwo,\gabste}), the lattice of D-brane charges is
not generated by these fractional branes alone. Indeed, it is manifest
from Table~2 that there is a boundary state in the untwisted R-R
sector that is invariant under all projection operators provided that
all $s_i$ are odd (\ie\ equal to $1$) and that $r$ is odd for IIA and
even for IIB. One can therefore construct an additional `bulk'
D-brane, \ie\ a boundary  state with only untwisted components
\refs{\stef}. However this bulk brane is {\it not} the minimally
charged object. Indeed, one can construct an `almost fractional'
boundary state whose charge is half the charge of the bulk brane. The
moduli space of the corresponding brane\footnote{$^\dagger$}{We
propose to call branes of this type {\it projective fractional}
D-branes.} consists of the different fixed planes of $g_1$, $g_2$ or
$g_3$. The boundary state description of the brane is slightly
different for the different branches of the moduli space, and we shall
in the following only give the explicit formula for the brane that
stretches between the fixed planes of $g_1$ defined by $x^6=x^8=0$,
where $x^5$ and $x^7$ take the values $0$, $\pi R^5$ and $0$, 
$\pi R^7$, respectively; the formulae for the other branches (and 
orientations) are analogous.    

Let us denote by ${\bf y}$ the position of the brane in the directions
that are unaffected by the orbifold action, by ${\bf a}$ the
coordinates in the $x^3,x^4$ directions on the fixed planes of $g_1$,
and by ${\bf b_i}, i=1,2,3,4$ the coordinates (in the
$x^5,x^6,x^7,x^8$ directions) of the four fixed planes. (So, for
example, ${\bf b_1}=(0,0,0,0)$, ${\bf b_2}=(\pi R^5,0,0,0)$, 
${\bf b_3}=(0,0,\pi R^7,0)$, ${\bf b_4}=(\pi R^5,0,\pi R^7,0)$.)
The relevant boundary state is then of the form 
\eqn\almost{\eqalign{|D(r;1,1,1);& {\bf y},{\bf a},\epsilon\rangle = 
|D(r;1,1,1);{\bf y},{\bf a}\rangle_{{\rm NS-NS;U}} 
+\epsilon 
  |D(r;1,1,1);{\bf y},{\bf a}\rangle_{{\rm R-R;U}}  \cr
& + \sum_{i=1}^{4} \left(|D(r;1,1,1);{\bf y},{\bf a}
          \rangle_{{\rm NS-NS;T}_{g_1, {\bf b_i}}}
+ \epsilon |D(r;1,1,1);{\bf y},{\bf a}
          \rangle_{{\rm R-R;T}_{g_1, {\bf b_i}}}\right)
\cr
& + |D(r;1,1,1);{\bf y},{\bf - a}\rangle_{{\rm NS-NS;U}} 
+\epsilon 
  |D(r;1,1,1);{\bf y},{\bf - a}\rangle_{{\rm R-R;U}}  \cr
& - \sum_{i=1}^{4} \left( |D(r;1,1,1);{\bf y},{\bf - a}
          \rangle_{{\rm NS-NS;T}_{g_1, {\bf b_i}}} 
+ \epsilon |D(r;1,1,1);{\bf y},{\bf - a}
          \rangle_{{\rm R-R;T}_{g_1, {\bf b_i}}}\right) \,,}}
where $r$ is odd for Type IIA and even for Type IIB, and 
$\epsilon=\pm 1$ distinguishes between the brane and the
antibrane. For simplicity, we have here described the brane without
Wilson line along $x^5$ and $x^7$ for which the signs of the
contributions of the four twisted sectors is the same. It follows from
the results of Table~2, together with the observation that both $g_2$
and $g_3$ act on ${\bf a}$ as ${\bf a}\mapsto-{\bf a}$ that the
boundary state in \almost\ is invariant under the action of the whole
orbifold group. The brane that is described by \almost\ can be thought
to consist of a brane at $({\bf y},{\bf a})$ together with an 
anti-brane at $({\bf y},{\bf -a})$ (where the untwisted R-R charges of
the two branes are the same, whereas the charges with respect to the
$g_1$-twisted sectors are opposite). The open string that corresponds 
to this boundary state has therefore a $2\times 2$ Chan-Paton matrix
\eqn\chan{\left(\matrix{ a&b\cr c&d}\right) \,,}
where $a$ ($d$) labels the string that begins and ends at the brane at 
${\bf a}$ (${\bf -a}$), whereas $b$ denotes the string that begins at
${\bf a}$ and ends at ${\bf -a}$, and $c$ denotes the same string with
the opposite orientation.  In terms of the open string, the orbifold
generators act on the Chan-Paton matrix by conjugation,           
\eqn\conj{\left(\matrix{ a&b\cr c&d}\right) \mapsto 
\gamma(g_i) \left(\matrix{ a&b\cr c&d} \right)\gamma(g_i)^{-1}\,.} 
We can read off from \almost\ how the Chan-Paton matrix transforms
under the three orbifold actions, and we find that
\eqn\trans{\eqalign{ g_1: \quad & \left(\matrix{ a&b\cr c&d}\right)
\mapsto  \left(\matrix{ a& -b\cr -c  &d}\right) \cr
g_2: \quad & \left(\matrix{ a&b\cr c&d}\right) \mapsto  
\left(\matrix{ d& \pm c\cr \pm b &a}\right) \cr
g_3: \quad & \left(\matrix{ a&b\cr c&d}\right) \mapsto  
\left(\matrix{ d& \mp c\cr \mp b &a}\right)\,.}}
For example, $g_1$ acts with a minus sign on the strings that run
between the brane and the anti-brane, and $g_2$ and $g_3$ exchange the
brane and the anti-brane. On the level of this discussion it is
impossible to fix the signs on the off-diagonal elements in the action
of $g_2$ and $g_3$, but consistency with the group relations, in
particular $g_2 g_2=e$, $g_3 g_3=e$ and $g_3=g_1 g_2$, determines the
relative signs as above. 

The matrices $\gamma(g_i)$ that implement the transformations
described in \trans\ by conjugation as in \conj\ are 
\eqn\reps{\eqalign{\gamma(g_1)& = 
\left( \matrix{1&0 \cr 0 & -1}\right) \cr
\gamma(g_2)& = 
\left( \matrix{0&\pm 1 \cr 1 & 0}\right) \cr
\gamma(g_3)& = 
\left( \matrix{0&\mp 1 \cr 1 & 0}\right)\,.}}
These matrices define a {\it projective} representation of
$\Zop_2\times\Zop_2$; indeed, if we consider the upper sign in 
\reps\ we have 
\eqn\projrep{ \gamma(g_i) \gamma(g_j) = c(g_i,g_j) \gamma(g_ig_j) \,,}
where $c$ is the cocycle that was defined in
\torc.\footnote{$^\ddagger$}{The lower sign in \reps\ leads to a
cocycle $c'$ that differs from $c$ by the coboundary $c_e=c_{g_1}=1$,
$c_{g_2}=c_{g_3}=+i$.} The existence of a brane with such properties
was predicted in \refs{\douglas,\dougfiol}; their argument was based
on the observation that the theory with discrete torsion has a brane
that leads to a projective representation of the orbifold group, and
that T-duality along $x^3,x^5,x^7$ maps the orbifold theory with
discrete torsion to the one without discrete torsion
\refs{\vafawitten}.

\subsec{The theory with discrete torsion}

For the theory with discrete torsion, \ie\ the theory where $g_i$ acts
as $\hat{g_i}$ on the twisted sectors, the analysis is completely
analogous. Fractional branes exist now for 
\eqn\fractionalt{ \hbox{Fractional D-branes}  \qquad \qquad
\eqalign{\hbox{\bf IIA:} \quad & \hbox{$r$ odd and $s_i$ odd,} \cr
\hbox{\bf IIB:} \quad & \hbox{$r$ even and $s_i$ odd.}}}
In addition, there are projective fractional branes (whose boundary
state is given by a similar expression as in \almost) for 
\eqn\projectivet{ \hbox{Projective fractional D-branes}  \qquad \qquad
\eqalign{\hbox{\bf IIA:} \quad & \hbox{$r$ even and $s_i$ even,} \cr
\hbox{\bf IIB:} \quad & \hbox{$r$ odd and $s_i$ even.}}}
The fractional branes are the states in the theory that are charged
under the massless fields from the twisted R-R sectors; if these
branes did not exist, the theory would not possess {\it any} states
that are charged under these fields. 

The projective fractional D-branes were first discussed (in the
uncompactified theory) in \refs{\douglas,\dougfiol}, where it was also
observed that their moduli space has three branches (at each of the
fixed planes) as we have found above. Under T-duality along $x^3,x^5$
and $x^7$, say, the IIA (IIB) theory without discrete torsion is
mapped to the IIB (IIA) with discrete torsion and vice versa. On
D-branes, this T-duality transformation leaves $r$ invariant, and
changes each $s_i$ by $\pm 1$. This is consistent with the D-brane
spectrum of the two theories that we have found above.

\newsec{Open-closed consistency condition}
\pano

It was shown by Gomis \refs{\gomis} that the representations of the
orbifold group that appear in the open string description are
constrained in terms of the actual representation of the orbifold
group on the various twisted sectors of the closed string theory; this
is a consequence of the open-closed consistency condition that was
first considered, in a slightly different context, in \refs{\polch}. 
Superficially, the analysis of Gomis seems to imply that for orbifolds
with discrete torsion {\it only} projective representations of the
orbifold group can occur in the open string; as we have seen above,
this would be in conflict with T-duality (and our discussion of
Dirichlet branes in these theories). We shall now explain that his
argument, correctly interpreted, is in precise accord with what we
have found above.

Following \refs{\gomis}, let us consider the disk diagram where we
insert a closed string state in the $g_i$-twisted sector at the
centre of the disk, and an open string vertex operator at the
boundary. As is explained in \refs{\gomis}, this amplitude is
proportional to 
\eqn\amp{ \hbox{Tr} (\gamma(g_i) \lambda) \;
   \langle V(\bphi,0) V(\psi,1)\rangle \,,}
where $\lambda$ denotes the Chan-Paton matrix of the open string field
$\psi$, and $\bphi$ is the state in the $g_i$-twisted sector. Here
$\gamma(g_i)$ arises because the state in the $g_i$-twisted sector
generates a branch cut from the centre of the disk to the boundary
along which fields jump by the action of $g_i$.  

The consistency condition on the allowed representations in the open
string arises from the constraint that this amplitude must be invariant
under the action of the orbifold group. As we have argued before, 
the action of the orbifold group on the closed string twisted sectors
is given by $\hat{g}$ as in \torsiontwo, where the action of $g$ on
the twisted sectors is defined in some natural way, and $\epsilon$
describes the relative discrete torsion with respect to the reference
theory.\footnote{$^\star$}{This is where our analysis differs from
that of Gomis: he assumes that the action of the orbifold group on the
twisted sectors is unmodified for the case of discrete torsion, and
that only the condition on physical states is modified as in
\twisted. These two points of view are equivalent for the closed
string sector of the theory, but they lead to different conclusions
once open strings are considered as well. As we will see, our point of
view reproduces the above D-brane spectrum that is consistent with
T-duality.} For general $\epsilon$, the condition that the amplitude
\amp\ is invariant under the action of the orbifold group therefore
becomes    
\eqn\ampeq{\eqalign{\hbox{Tr} (\gamma(g_i) \lambda) \;
   \langle V(\bphi,0) V(\psi,1)\rangle 
& = \hbox{Tr} (\gamma(g_i) \gamma(g_j) \lambda \gamma(g_j)^{-1})  \;
   \langle V(\hat{g_j} \bphi,0) V(g_j \psi,1)\rangle \cr
& = \epsilon(g_j,g_i) 
    \hbox{Tr} (\gamma(g_i) \gamma(g_j) \lambda \gamma(g_j)^{-1}) \;
   \langle V(g_j \bphi,0) V(g_j \psi,1)\rangle\,,}}
where we have used that the action on the open string Chan-Paton
indices is defined by \conj.  

Let us consider an open string state $\psi$ that is invariant under
the action of the orbifold group, $g_j \psi=\psi$, and let us
denote by $\delta_j(\bphi)$ the eigenvalue of $\bphi$ under the action
of $g_j$, $g_j \bphi=\delta_j(\bphi)\bphi$. If $\bphi$ is a physical
state in the orbifold theory with relative discrete torsion
$\epsilon$, then we have to have
$\delta_j(\bphi)\epsilon(g_j,g_i)=+1$. If in addition the amplitude 
$\langle V(\bphi,0) V(\psi,1)\rangle$ does not vanish, then the
consistency condition implies that  
\eqn\consist{ \hbox{Tr} (\gamma(g_i) \lambda) 
= \hbox{Tr} (\gamma(g_i) \gamma(g_j)\lambda \gamma(g_j)^{-1}) \,.}
The Chan-Paton matrix $\lambda$ is arbitrary, and this statement is
therefore equivalent to 
\eqn\consistonen{ \gamma(g_j) \gamma(g_i) = \gamma(g_i) \gamma(g_j)\,.} 
In particular, it then follows that the representation of the orbifold
group defined by $\gamma$ is a proper (not a projective)
representation. This conclusion applies to those open strings that
have a non-vanishing coupling with physical states in the twisted
sector of the orbifold theory. This is in particular the case for the 
fractional D-branes we have discussed above.

On the other hand, the situation that was considered by Gomis
corresponds to the case when $\delta_j(\bphi)=+1$, \ie\ when the
open string state couples to a closed string state that is physical in
the theory without discrete torsion, but unphysical provided that
$\epsilon(g_i,g_j)\ne 1$. Again under the assumption that the
corresponding overlap does not vanish, an analogous argument then
implies that  
\eqn\consistonep{ \gamma(g_j) \gamma(g_i) = \epsilon(g_j,g_i) 
\gamma(g_i) \gamma(g_j)\,.} 
If we rewrite $\epsilon(g_j,g_i)$ in terms of the cocylce $c$ as in 
\eps, this becomes
\eqn\consisttwop{ \gamma(g_j) \gamma(g_i) c(g_i,g_j)
= c(g_j,g_i) \; \gamma(g_i) \gamma(g_j)\,.}
The representation of the orbifold group in the open string is then
the projective representation described by 
\eqn\projreps{ \gamma(g_i) \gamma(g_j) = c(g_i,g_j) \gamma(g_i g_j)
\,.} 
This analysis applies to the projective fractional branes for which
the twisted closed string states to which the D-brane would normally
couple are unphysical.

{}From this point of view, the question of whether the representation
of the orbifold group on the open string Chan-Paton indices is a
proper representation or a projective representation does {\it not}
depend on whether the theory in question is an orbifold with or
without discrete torsion; it only depends on the transformation
properties of the twisted sector states to which the open string state
couples.

\newsec{Conclusions}
\pano

In this paper we have re-examined the D-brane spectrum of a certain
$\Zop_2\times\Zop_2$ orbifold with and without discrete torsion. We
have argued, on general grounds, that the analysis for the two cases
must be analogous, and we have shown that this leads to a D-brane
spectrum that is consistent with T-duality that relates the theory with
and without discrete torsion \refs{\vafawitten}. The picture that
seems to be emerging is that the representation of the orbifold group
in the open string description can either be a proper or a projective
representation, irrespective of whether the orbifold theory has
discrete torsion or not. The emergence of projective representations
of the orbifold group is merely related to discrete torsion in the
sense that only orbifold theories that admit discrete torsion also
admit projective representations that are not equivalent to proper
representations.\footnote{$^\dagger$}{Every such projective
representation gives rise to a non-trivial co-cycle in
$H^2(\Gamma,U(1))$.} This conclusion is somewhat different from what
was argued for in \refs{\douglas,\dougfiol,\gomis}.   
We have also given a boundary state description for branes that lead
to projective open string representations, and we have found that
their moduli space has the same structure as perdicted in
\refs{\douglas,\dougfiol}.

In this paper we have only analysed the $\Zop_2\times\Zop_2$ case that
is special in that T-duality relates the theory with discrete torsion
to the one without. It would be interesting to see how the findings of
this paper generalise for more general orbifolds with (and without)
discrete torsion; this is currently under consideration
\refs{\cragab}. It would also be interesting to understand whether
non-BPS D-branes can also have projective representations of the
orbifold group, and how this fits together with the various decay
processes of non-BPS branes into brane anti-brane pairs.

\vskip 1cm

\centerline{{\bf Acknowledgements}}\pano

I thank Bogdan Stefa\'nski for drawing my attention to this problem,
and for useful conversations. I am grateful to Andy Neitzke for
useful conversations about group cohomology. I also thank Michael
Douglas and Jaume Gomis for helpful correspondences.   

I am grateful to the Royal Society for a University Research
Fellowship. I also acknowledge partial support from the PPARC SPG
programme, ``String Theory and Realistic Field Theory'',
PPA/G/S/1998/00613. 
\vskip 1cm

\appendix{A}{The construction of the two-cocycle}

Let us first determine the most general set of phases $\epsilon(g,h)$
that satisfy \consistone\ and \consisttwo. Recall that every finitely
generated abelian group $\Gamma$ can be written as (see for example
\refs{\vafa}) 
\eqn\group{ \Gamma = \Zop_{m_1} \times \Zop_{m_2} \times \cdots \times
\Zop_{m_k} \,,}
where $m_i$ is a factor of $m_{i+1}$. Let us denote by $\alpha_i$ the
generator of $\Zop_{m_i}$. Then every set of discrete torsion phases
is uniquely determined by the set of phases
\eqn\sol{ \epsilon_{ij} = \epsilon(\alpha_i,\alpha_j) \quad
\hbox{where $i<j$.}}
Indeed, it is easy to check that \consistone\ and \consisttwo\ imply
\pro, and the second equation in \pro\ determines then $\epsilon$ on
all pairs of generators. Using the representation property
\consisttwo\ this finally fixes $\epsilon$ for all pairs
$(g,h)$. Provided that $\epsilon_{ij}$ is a $m_i$th root of unity, 
this construction is well-defined. Since the resulting phases satisfy
by construction \consisttwo\ and \pro, it follows by the same
arguments as in the main part of the paper that they also satisfy
\consistone. We have therefore shown that the set of possible discrete
torsion phases is given by 
\eqn\poss{ \Zop_{m_1}^{k-1} \times \Zop_{m_2}^{k-2} \times \cdots
\times \Zop_{m_{k-1}}\,.}
The set of possible discrete torsion phases is actually a group since
the product of two sets of discrete torsion phases defines another set
of discrete torsion phases. This group is generated by the 
{\it primitive} discrete torsion phases for which $\epsilon_{ij}=1$
for all but one pair $i<j$ for which $\epsilon_{ij}$ is a primitive
$m_i$th root of unity.  

The set of cocylces is also an abelian group (where the group
multiplication is also given by pointwise multiplication). In order to
construct a cocycle $c$ for each set of discrete torsion phases
$\epsilon$ (so that $\epsilon$ is determined in terms of $c$ by \eps),
it is therefore sufficient to construct such a cocycle for the 
primitive discrete torsion phases only. This can be done indirectly,
by constructing a certain projective representation of $\Gamma$
\refs{\dougfiol}.  

In order to simplify notation, let us consider the case where the
primite discrete torsion phases are defined by $\epsilon_{12}\ne 1$,
with $\epsilon\equiv\epsilon_{12}$ a primitive $m\equiv m_1$th root of
unity. There are two cases to consider: if $m$ is odd, we construct a
$m$-dimensional projective representation by  
\eqn\repp{
\gamma(\alpha_1) = 
\pmatrix {&0 &1 &0 &\cdots &0 \cr
          &0 &0 &1 &\cdots &0 \cr
          &\cdots &\cdots &\cdots &\cdots &\cdots \cr
          &0 &0  &\cdots &0 &1 \cr
          &1 &0  &0  &\cdots &0 }\qquad
\gamma(\alpha_2) = 
\pmatrix {&0 &\epsilon &0 &\cdots &0 \cr
          &0 &0 &\epsilon ^2 &\cdots &0 \cr
          &\cdots &\cdots &\cdots &\cdots &\cdots \cr
          &0 &0  &\cdots &0 &\epsilon^{m-1} \cr
          &1 &0  &0  &\cdots &0}\,,}
together with $\gamma(\alpha_j)={\bf 1}_{m}$. On the other hand, if
$m$ is even, $\gamma(\alpha_j)$ for $j\ne 2$ is as above, and we
define 
\eqn\repe{
\gamma(\alpha_2) = \pmatrix {&0 &\delta &0 &\cdots &0 \cr
          &0 &0 &\delta ^3 &\cdots &0 \cr
          &\cdots &\cdots &\cdots &\cdots &\cdots \cr
          &0 &0  &\cdots &0 &\delta^{2m-3} \cr
          &\delta^{2m-1} &0  &0  &\cdots &0}\,, \qquad 
          \hbox{($m$ even)}} 
where $\delta^2=\epsilon$. This construction guarantees that
$\gamma(\alpha_j)^m={\bf 1}$ for all $j$. Furthermore we have
\eqn\comm{ \gamma(\alpha_1)  \gamma(\alpha_2) 
               = \epsilon \gamma(\alpha_2)  \gamma(\alpha_1) \,,}
while all other generators commute pairwise. 

Next we extend this to a (projective) representation of $\Gamma$ by
defining 
\eqn\ext{ \gamma(g) = \prod_{j} \gamma(\alpha_{i_j}) \,,}
where we choose for each element $g\in\Gamma$ a realisation as 
$g = \prod_{j} \alpha_{i_j}$, and we pick a specific order for the
$\gamma(\alpha_{i_j})$ on the right-hand-side of \ext. Because of the
commutation relations and the property that 
$\gamma(\alpha_j)^m={\bf 1}$ for all $j$, we then find that
\eqn\projrep{ \gamma(g) \gamma(h) = c(g,h) \gamma(gh) \,,}
where $c(g,h)$ are certain phases. These phases satisfy the cocycle
condition \cocycle\ since the representation of $\Gamma$ is
associative. 

It follows directly from \comm\ that $\epsilon(g,h)$, defined by \eps,
satisfies $\epsilon(\alpha_i,\alpha_j)=1$ unless $(i,j)=(1,2)$ or
$(i,j)=(2,1)$, and that $\epsilon(\alpha_1,\alpha_2)=\epsilon$. Since
each cocylce defines a set of phases that satisfy \consistone\ and
\consisttwo, it therefore follows that the above cocycle reproduces
indeed the desired discrete torsion phases.
\vskip 0.2cm

Finally, in order to prove that the correspondence between discrete
torsion phases and cocycles is one-to-one, it remains to check that
the map defined by \eps\ is injective. Let us therefore assume that
for a given cocycle $c(g,h)$, all $\epsilon(g,h)=+1$, \ie\ that
$c(g,h)=c(h,g)$ for all $g,h\in\Gamma$. We want to show that we can
find a coboundary so that $c'$, as defined in \cobound, satisfies
$c'(g,h)=1$ for all $g,h\in\Gamma$.

Without loss of generality (by choosing $c_{e}=c(e,e)^{-1}$) we may
assume that $c(e,e)=1$. The cocycle condition \cocycle\ then implies
that $c(e,g)=c(g,e)=1$ for all $g\in\Gamma$. Let us then define
$c_g$ for all $g\in\Gamma$ in terms of $c_{\alpha_{i}}$ by
the formula 
\eqn\define{c_{\alpha_{i_1}\cdots\alpha_{i_l}} = 
c(\alpha_{i_1}\cdots\alpha_{i_{l-1}},\alpha_{i_l})\,
c(\alpha_{i_1}\cdots\alpha_{i_{l-2}},\alpha_{i_{l-1}}) \cdots\,
c(\alpha_{i_1}\alpha_{i_2},\alpha_{i_3})\,
c(\alpha_{i_1},\alpha_{i_2}) \,
\prod_{j=1}^{l} c_{\alpha_{i_j}} \,.} 
Using the cocycle condition \cocycle\ and the property that
$c(g,h)=c(h,g)$ it is easy to see that this definition is independent
of the order of the $\alpha_i$. In order to show that it respects the
group relations, we observe that 
\eqn\property{c_{\alpha_{i_1}\cdots\alpha_{i_l}
\alpha_{j_1}\cdots\alpha_{j_m}}
= c(\alpha_{i_1}\cdots\alpha_{i_l},\alpha_{j_1}\cdots\alpha_{j_m})
c_{\alpha_{i_1}\cdots\alpha_{i_l}}
c_{\alpha_{j_1}\cdots\alpha_{j_m}}\,,} 
where we have again used the cocycle condition. We can choose
$c_{\alpha_{i}}$ so that $c_{\alpha_i \cdots \alpha_i}$, as defined by
the right-hand-side of \define, satisfies
\eqn\test{c_{\underbrace {\alpha_i\cdots\alpha_i}_{m_i}}=c_{e}=1\,.} 
Because of \property\ and the fact that $c(e,g)=c(g,e)=1$, this then
implies that \define\ respects the group relations. Finally, it is
manifest from \property\ that $c'(g,h)$, defined by \cobound, 
satisfies then $c'(g,h)=1$ for all $g,h\in\Gamma$.

\listrefs

\bye